# Experimental measurements of a joint 5G-VLC communication for future vehicular networks

Dania Marabissi [1], Lorenzo Mucchi [1]*, Stefano Caputo [1], Francesca Nizzi [1], Tommaso Pecorella [1], Romano Fantacci [1], Tassadaq Nawaz [3], Marco Seminara [3,4], and Jacopo Catani [4,3]

[1] Dept. of Information Engineering, University of Florence, Italy; {name.surname}@unifi.it
[2] Dept. of Physics and Astronomy, University of Florence, Italy; {name.surname}@unifi.it
[3] European laboratory of non linear spectroscopy (LENS); {surname}@lens.unifi.it
[4] National Institute of Optics (INO-CNR); {name.surname}@ino.cnr.it
* Correspondence: lorenzo.mucchi@unifi.it; Tel.: +39-055-275-8539 (F.L.)

Abstract: One of the main revolutionary features of 5G networks is the ultra-low latency that will enable new services such as those for the future smart vehicles. The 5G technology will be able to support extreme-low latency. Thanks to new technologies and the wide flexible architecture that integrates new spectra and access technologies. In particular, Visible Light Communication (VLC) is envisaged as a very promising technology for vehicular communications, since the information can flow by using the lights (as traffic-lights and car lights). This paper describes one of the first experiments on the joint use of 5G and VLC networks to provide real-time information to cars. The applications span from road safety to emergency alarm.

## 1. Introduction

In last decades the continuous increase of capacity demand resulting from massive data growth has asked for rapid changes of wireless communication networks. Nowadays, this evolutionary trend is moving toward the fifth generation (5G) of wireless systems. 5G is envisioned to have a significant technology gap compared to previous cellular networks: very high data rate, extremely low latency, high cell capacity, massive number of connected devices, guaranteeing energy and cost-efficiency. To face with this challenges, 5G networks are expected to deploy a high number of cells and to include additional spectrum respect to current systems. Consequently, communications in millimeter-wave (mmWave) spectrum, with short-range and large bandwidth availability, are considered a key 5G-technology. Also, other spectra and technologies are under investigation. In particular, Visible Light Communication (VLC) is considered a promising complementary technology to mmWave for short-range communication scenarios, especially for indoor and hot-spot connections [1]. VLCs can provide very high data-rates, low-energy consumption, low latency and low implementation costs. IEEE has standardized the physical layer (PHY) and medium access control (MAC) sublayer for short-range optical wireless communications (OWC), including VLC and optical camera communications (OCC) [2]. Consequently, while the 5G standardization process is ongoing, the integration of VLC in 5G systems is under investigation [3-6]. In [3] and [4], hypotheses on the integration between 5G and VLC are envisioned. In [5], the integration of VLC segment with 5G backhaul is investigated, while [6] and [7] give an overall general description of the benefits that the VLC can bring to 5G networks.

Apart the research and standardization processes, launching a new technology requires field trials to define and test the key performance indicators (KPIs) as well as to determine how the users can maximize the exploitation of the new capabilities of the 5G network. Different field trials and tests have been carried out all around the world and are currently on going. Anyway, all these tests are mainly focused on the 5G New Radio (NR) technologies: massive Multiple Input Multiple Output (mMIMO), 3D-beamforming, carrier aggregation, use of new air interfaces, and advanced solutions for the core network [8], [9]. Differently, the integration of 5G NR and VLC is not largely tested, only few papers present experimental results. In [10], a performance evaluation is presented where the 5G NR air frame is adapted to the VLC transmission, while in [11] resource allocation in VLC networks is discussed.

This paper presents a significant contribute in this context. In fact, it focuses on a field-test that integrates 5G communication capabilities with VLCs in a vehicular scenario. The aim is mainly to evaluate performance that can be actually achieved. To the best of our knowledge this is the first case presenting an effective VLC_5G integrated field-trial in this application scenario. The main goal is to test the end-to-end latency of the communication that can be offered by this integrated network. The Italian Ministry of Economical Development (MiSE) started a 5G field-trial project at the end of 2017 in the cities of Prato e L'Aquila. The entire experimental project lasts 4 years and includes a pre-commercial experimentation during the last year. Two telecommunications operators, WindTre and OpenFiber, are the coordinators of the project since they both owwnn significant network infrastructures both wireless and optical fibers. This paper first describes the main characteristics of the field-trial environment and then details the experiment carried out to deliver road information to vehicles. Information collected by sensors connected to the 5G network, is distributed by the traffic light with near-zero latency thanks to the VLC. This paper is an extension of two previous papers [12] [13]. In particular, [12] presents a general description of the 5G project framework that is here recalled only on its main aspects, while [13] presents the description of the smart mobility use case. This paper focuses on the same use case but providing KPIs measurements and analysis of the performance of the integration between the 5G network and a VLC-based V2I communication system, achieved at the end of the 5G project. In addition, we have included the distribution of the empirical data of the latency for the 5G link and for the overall system (5G link and VLC link), as well as the best fitting distribution, derived by using the Bayesian information criterion (BIC). The probability density function (PDF) of the empirical data and fitting PDFs are reported. Also, cumulative distribution function (CDF) and CDF error functions are shown in the paper.

The rest of the paper is organized as follows. Sec. 2 describes the overall system model of the integration of the VLC system into the 5G network. Sec. 3 briefly highlights the 5G testing network, the use cases and the field trials. Sec. 4 shows the details of the VLC system, while Sec. 5 reports and discusses the experimental results. Sec. 6 concludes the paper.

2. VLC-5G Integration System Model

Before describing the 5G field trial, in this section we present the system model of the VLC-5G integration used in the field-trial that will be detailed later. The model is presented in Fig. 1. In particular, the VLC segment is provided by the traffic lights of a (smart) city, while the 5G network provides the connectivity to the traffic lights.

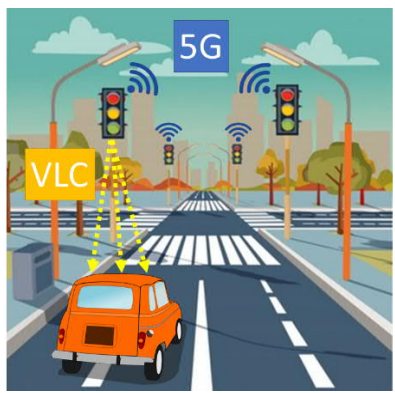

Figure 1. VLC-5G network topology.

Let us assume that a set {1, ..., s, ..., S} of traffic lights are serving a set {1, ..., u, ..., U} of vehicular users. Each traffic light is connected to 5G network and it is capable to generate a VLC signal towards the incoming vehicles. Each traffic light is assumed to be within the coverage of a 5G cell. Vehicles are assumed to be equipped with a VLC receiver. Since each traffic light has a directive beam, we assume that traffic lights do not interfere each other, i.e., the VLC bandwidth can be reused.

The signal-to-noise ratio (SNR) for the uth user served by the sth traffic light through a VLC link can be written as [14]

$$\gamma_{u,s} = \frac{R_{PD} H_{u,s} P_s}{\xi} \quad (1)$$

where $R_{PD}$ denotes the responsivity of the photodiode, $H_{u,s}$ is the line-of-sight (LOS) VLC channel gain between the sth traffic light and uth vehicle, $P_s$ is the transmission power, and $\xi$ denotes the cumulative noise power. The channel gain $H_{u,s}$ depends on several variables: photodetector area, angle of irradiance, angle of incidence, signal transmission coefficient of the optical filter, refractive index of the optical concentrator, user's field of view (FOV), and the distance d between the traffic light and the user vehicle.

## 3. 5G Field Trials

### 3.1 Field-trials organization

The 5G field-trial project has been organized in phases and follows the standardization process. The 3rd Generation Partnership Project (3GPP) has completed the first version of the 5G standard ready for deployment in 2017. It is named Non-Standalone 5G NR (NSA 5G NR) and it requires fall back to Long Term Evolution (LTE) networks for partial operation. Currently, the Stand Alone (SA) 5G network architecture is under definition (to be completed in June 2020).

Project organization:
1) Set-up phase: the ZTE Research and Innovation Lab has investigated and tested new 5G technologies.
2) Roll-out phase: several NR base stations (BSs) have been deployed to test all network elements and basic network functions. Initially, the deployment relays on the existing LTE-Core Network (CN) following the NSA network architecture. At the end of the project the SA network architecture (with a 5G-CN) will be deployed and evaluated.
3) Service phase: extensive field trials to validate the KPIs and to test innovative services provided on the 5G network infrastructure.

### 3.2 5G Network

#### 3.2.1 Network Architecture

The 5G network architecture deployed for the testbed is based on a network slicing approach.

Network slicing is considered one of the pillars of 5G systems. Specific functions can be designed to create and manage dedicated end-to-end logical networks, without losing the economies of scale of a common physical infrastructure. Each logical network is tailored to provide a specific service and/or provide a particular tenant with a given level of guaranteed network resources [15]. Consequently, 5G systems can support a wide variety of vertical markets that originate a wide range of services.

The 5G network architecture considered in the field-test has a logical organization on three-layers (Figure 2). These are used for diversification of the service layer from the network functions and the physical infrastructure. Physical infrastructure is the first layer, whose main objective is the managing of the physical resources. Network functions is the second layer, whose scope is functions' customization. Service is the third layer: it maps the Service Level Agreements (SLAs), Quality of Service (QoS), and required functionality into the slice's configurations. An orchestrator manages the three layers by mapping the resources available at different layers to the slices.

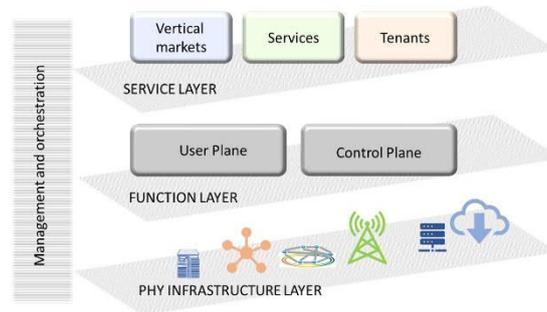

Figure 2. Network slicing conceptual architecture.

3.2.2 Physical Infrastructure

In the deployed infrastructure the Cloud-Radio Access Network (RAN) approach has been adopted. BSs are composed by several radio frequency elements called (active Antenna Units – AAU) and the Base Band Unit (BBU) that is the element of processing. The BBU can ben centralized in a single point or virtualized in the cloud and represents the smart-element of the BS while only simple RF equipment are needed at the network edge.

AAUs are able to work on different frequencies, thus a multimode network is deployed. In particular, a heterogeneous multi-layer 5G cellular architecture is considered where cells of different coverage areas are overlapped and provide services in different frequency bands (e.g., 3.7 GHz and mmWave) and with different access technologies (e.g., 5G-NR, WiFi and VLC). In this case, a dual-connectivity approach is adopted. The 5G-NR AAU operating in low frequency bands provides basic services on a wide area, while small cells operating on mmWave and visible light spectra provide high data-rate services indoor and in hot-spots.

AAUs and BBU are connected though the fronthaul link that is characterized by low-latency and high-speed thanks to the adoption of the Common Public Radio Interface (CPRI) [16]. CPRI is an interface that defines the transmission of digital-radio over fiber and allows a transmission of data with a fixed bit rate over a dedicated channel.

A single BBU and a single AAU connected to a channel emulator and to a basic User Equipment (UE) were used during the very initial tests. Successively, several BBUs and AAUs, connected to the LTE-CN following the 3GPP NSA deployment scenario Option 3 and 3a [17], have been deployed in the two cities. In particular, the LTE radio and CN were used as an anchor for mobility management and coverage, while adding new 5G carrier. Hence, the LTE-RAN connects the LTE-CN with the 5G NR.

The FlexE [18] standard was used to manage via software the transport network. This allowed a flexible reconfiguration of the network, making the physical layer transparent to the service layer. 100G Ethernet rings provided connections to BBUs, while optical links provided connections with the core network.

NSA deployment allows to validate KPIs mainly related to control plane and user plane latency, user and cells peak data rates, and to test network slicing approach [15].

The SA 5G NR architecture will be deployed during the last part of the roll-out phase. In particular, 3GPP Option 2 [19], i.e., 5G NR devices are directly connected to the new 5G CN, is considered in the deployment of the network. This solution is independent on 4G network deployments and it provides simpler implementation. Anyway, the above solution requests the 5G end-to-end network to be completely defined before the pre-commercial phase, and substantial investments to provide the service coverage over all the territory. In this phase KPIs related to mobility and handover will be tested.

The 5G NR improved air interface is based on:
- operations in multiple frequency bands;
- mMIMO techniques;
- non-orthogonal multiple access;
- dense network deployment.

### 3.3 Use Cases

The project aims to validate the 5G and its role of new digital economy creators. In fact, 5G is does not provide only the enhancement of current wireless systems, but it also represents a framework where new services and applications will be provided. Consequently, 5G can support a different vertical market that need of a wide range of services each one characterized by specific requirements [19].

In this context, the project defines several use cases to test the provisioning on the 5G network of innovative services that are under deployment.

A brief description is provided in Table 1, while the use case on Smart Mobility is detailed in the next section.

Table 1. 5G project Use Cases.

| Use Case | Description |
| --- | --- |
| *e-Health* | A platform that provides personalized care and assistance with guaranteed quality of service and continuity for telemedicine, telemonitoring and analysis of behavioural habits. |
| *Smart industry* | A digital platform to provide Industry 4.0 services for the optimization of production processes, energy efficiency, maintenance and operation. |
| *Smart grid* | A management architecture inspired by blockchain protocol enabling new services and management methods of the load and generation assets |
| *IoT and sensors* | Connected sensors (following the *IoT paradigm*) for real-time remote control of the industrial processes, heavy machinery in hazardous environments, logistics optimization and products tracking. |
| *Structural health monitoring* | A monitoring service for buildings/infrastructures, reporting any anomaly of the most significant structural parameters even in emergency (e.g., earthquake) by means the use of sensors and drones |
| *Virtual reality for cultural heritage* | An immersive virtual visit of different type of cultural heritage with digital contents delivery by using the virtual reality and the augmented reality. |
| *Agriculture 2.0* | Support and improvement of the *Made in Italy* brand. Tracking of products and production processes in the Agro-Food sector. |

More details on the overall 5G testbed can be found in [12].

### 3.3.1 Smart Mobility Use Case

This section focuses on Smart Mobility use case that is the one for which we present the testbed and experimental results.

It is a widely shared policy perspective that smart mobility will address numerous social challenges about freight and passenger transport: safety, efficiency, energy saving, reliability, etc. In this context, 5G is envisaged as the ultimate technology for Smart Mobility, since features of Cellular-Vehicle-to-Everything (C-V2X) will be included as part of the cellular chipsets embedded into vehicles for their Vehicle-to-Network (V2N) communications. The vertical domain of smart mobility is therefore one of the primary driving sectors for progression towards 5G, and one of the most important use cases in the 5G project.

The Smart Mobility platform has two main goals:
1) Road monitoring: electric parking and charging points are being deployed for the purpose of monitoring the state of the road surface (presence of gaps, slope, traffic conditions, etc.) during regular everyday activities by installing in the vehicles a blackbox containing a 5G module for real-time transmission of information to a data processing center. Electric cars are equipped with a differential GPS that can map the geographic positions of the holes found during regular vehicle use with a precision of cm.
2) Advanced viability: vehicles share data with other vehicles and with a control center where data traffic information is smartly combined other information such as the city's temperature, the road status and other sensors information. The goal is to use real-time information for increasing car and driver health, comfort, and style of driving and for minimizing road traffic, congestion and consequent emissions.

The Smart Mobility platform is based on some key technologies:
a) Network Slicing. The 5G network is able to provide specific network slice for V2X communications in order to manage its own features independently on the other services. However, how slices can efficiently share the resources is still a challenging issue. The studying of practical algorithms is ongoing considering both the computational complexity and the ability to reconfigure the resources allocation following the variability of the vehicular network topology. In particular, one of the main challenges of the infrastructure layer is the virtualization and division of the RAN into slices due to spectrum limitation. In addition, the coexistence communications with the network (V2N) and among vehicles requires a high flexibility and dynamicity of the RAN.
b) MEC. Reduced network congestion and improved applications performance can be obtained by using the Multi-Access Edge Computing (MEC) paradigm, which introduces cloud-computing capabilities closer to the end-user within the access network. Data generated from vehicles and infrastructure can be efficiently processed by the MEC thus delivering locally-relevant contents to support smart driving services. The MEC allows ultra-low latency, high bandwidth and real-time access to the access network that can be leveraged by the applications.
c) Access point densification. Network capacity can be improved by deploying a large number of small cells in addition to traditional macrocells. Morevoer, in case of emergency or network unavailability, vehicles themselves could complement the public network becoming moving cells. Anyway, in case of coexistence of multiple cell-layers, a careful investigation on resource usage is required as well as on coordination strategies among all the cells.
d) Multi RATs. In the smart mobility paradigm, multiple radio access technologies (Multi-RAT) can be integrated into vehicles, which become a powerful mobile gateway. Both V2V and V2N communications (e.g., 802.11p, LTE, C-V2X, 5G, VLC) could ask for multi-RATs integration, although an accurate managing should be done for the exploitation of benefits and limitation of their drawbacks.

In this context, one of the most original experimentations is the integration of VLC in the proposed platform for V2N and V2V communications. A measurement campaign on the integration of 5G system and VLC for vehicular communications have been carried out in the project, and it is described in the following section.

4. VLC for Vehicular Services

Vehicular networks applications are envisioned to benefit of communication opportunity given by visible light [20] [21]. Visible light communications (VLC) show advantages which are not inherent in RF-based technology, e.g., huge unlicensed spectrum, robustness versus jamming, and much less interference. VLC is also a green technology since the same energy used to light or for road signalling can be used to communicate with vehicles.

A preliminary experiment has been conducted at the University of Florence for I2V communications which involves VLC (Figure 3) signals. A real traffic-light, located in a real urban road, was used as VLC transmitter. A VLC receiver was moved over a grid of points, at different distances, in front of the traffic light. The experiments provided a 200 kbps of data rate up to 40 m. The prototype included a custom LED driver installed in a regular traffic light. The digital information was inserted into the light of the red lamp by the LED driver which modulated the light based on the source information bits. In particular, the intensity of the LED lamp moves from +A (bit 1) to -A (bit 0) around a mean value, which is the nominal intensity of the traffic-light. The lens inside the traffic-light is a regular traffic-light lens. The traffic-light was located in an urban road in the industrial part of the city. The receiver is composed by a photodiode and a lens collecting the light. A high precision oscilloscope was connected to the VLC receiver for the display and record of the incoming VLC signal. The VLC signal was recorded for 400 µs in each point of a virtual grid from 2 to 40 m along the lanes of the road. These measurements and the subsequent analysis led us to develop an accurate propagation model as well as a performance evaluation [22][23].

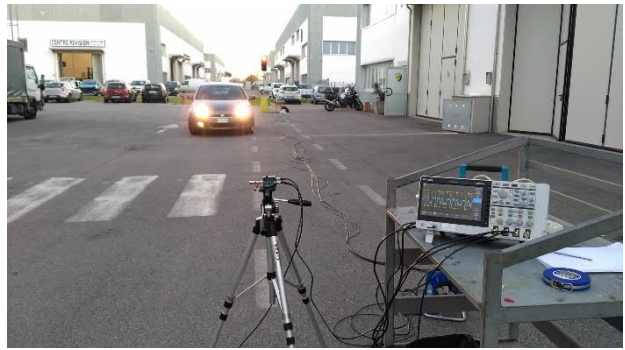

Figure 3. I2V measurements campaign using a VLC-based traffic-light (city of Prato).

As depicted in Figure 1, the framework envisioned to implement a 5G modem into each traffic light of the (smart) city, while each vehicle should include a VLC receiver. Computational complexity mainly depends on the type of information provided by the infrastructure to the incoming vehicles. As in large vehicular networks, information sent to vehicles or exchanged by vehicles is typically not "heavy", also due to the dynamic topology of the network. Messages to vehicles do not usually require a large amount of bits (alarms, traffic information, etc.), thus the computational complexity or the scalability of the network does not seem to be a hard task.

## 5. Advanced Viability experimental activity

### 5.1 Test-bed description

This section describes the experimental set up and results of a joint 5G-VLC networks for vehicular applications. In particular, road safety as well as emergency alarm services have been tested. The testbed was implemented at the Polo Universitario Città di Prato (PIN). The experimental set-up scheme is reported in Figure 4.

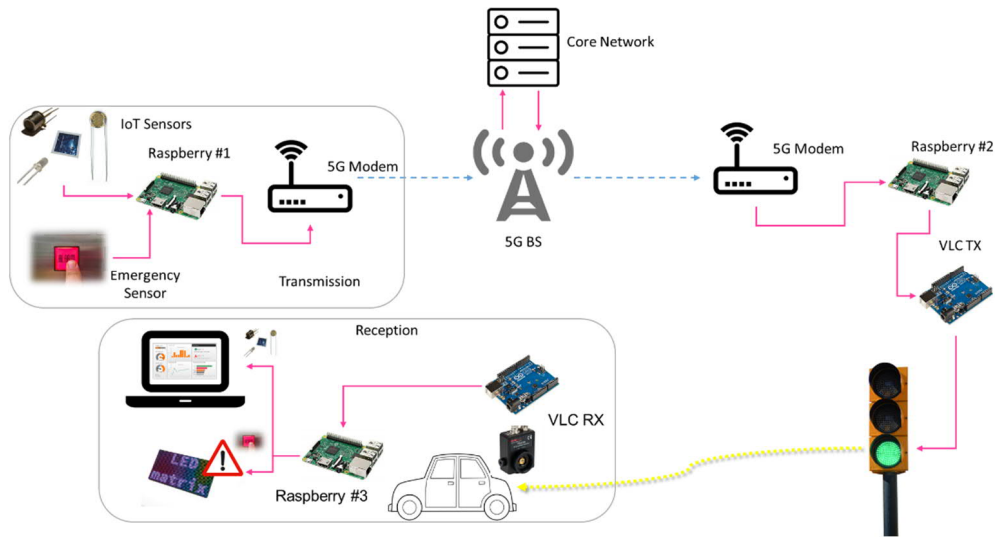

Figure 4. Test bed architecture of the 5G-VLC joint network.

In particular, the system was composed by three Raspberry Pi 3 Model B+, 3 WeMos D1 mini, an Italian regulation compliant traffic light, a VLC transmitter, a VLC receiver, and several different IoT sensors:
- Flame sensor: to simulate a fire alarm;
- Gyroscope/accelerometer: to detect an incident between to (scale model) cars;
- Temperature, humidity and pressure sensors: to detect the presence of ice on the road.

The first two Raspberries are used to collect data coming from the sensors and forward, through the 5G network, to the VLC network. The third Raspberry was used only to display in the vehicle the sensors values as well as the alarm. It is important to point out that the temperature/humidity/pressure sensors send the data periodically, while the other sensors send an alarm only if an event is triggered. The Raspberries are connected to the 5G Customer Premise Equipment (CPE), provided by the 5G network operator, though Ethernet Gigabit ports. The traffic-light and the car were connected using custom-designed TX and RX stages based on open source Arduino device [22]. The second Raspberry and the VLC block are connected by a serial cable (USB). The connection between the sensors and the raspberry is through the air: Raspberry provides a Wi-Fi Access Point and all the WeMos have an IP address. All the code in the Raspberries is written in Python 3 and all the code used to drive the sensor - WeMos is written in C/C++. The VLC transmitter and receiver are IEEE802.15.7-compliant and they have been specifically designed/prototyped for I2V communications (see [22] for all hardware implementation details). The implementation design of the transmitter includes on-off keying (OOK) modulation with Manchester encoding, while the receiver collects the light by using an aspherical 2" uncoated lens and it is designed to reject the sun light. The most expensive part is the photodiode. Although the initial cost of the VLC prototype could be considered high (few hundreds of euros), after industrialization the cost could be reduced to few tens of euros.

5.2 Experimental Results

The performance of the overall joint network is measured by the latency metric. The results are taken by an oscilloscope connected at the first Raspberry (yellow line in Figures 5 and 6), at the second Raspberry (purple line) and at the VLC receiver (green line). The yellow line, when high, represents an alarm trigger; the rising edge in the pink line represents the time needed to forward the packet on the USB cable; the green line represents the packet arrived at the VLC receiver.

Figures 5 and 6 show the maximum and the minimum measured end-to-end latency time. With end-to-end we intend here the time interval from the generation of one packet by the sensors to the correct reception of the packet by the VLC receiver in the car. The rectangle named A in Figures 5 and 6 shows the 5G latency time, while the rectangle named B reports the VLC latency time. The portion between the two rectangles represents the processing time introduced by the third Raspberry Pi3.

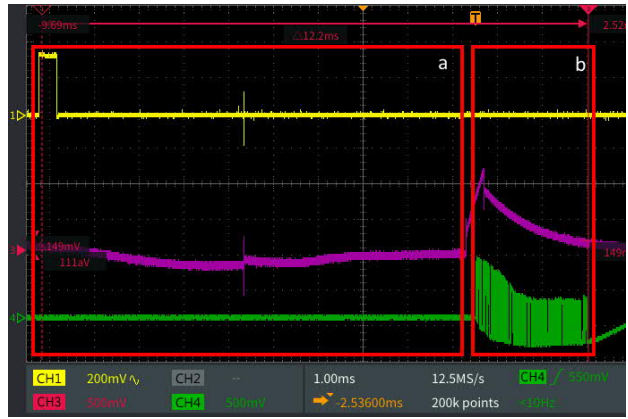

Figure 5. Maximum measured end-to-end latency.

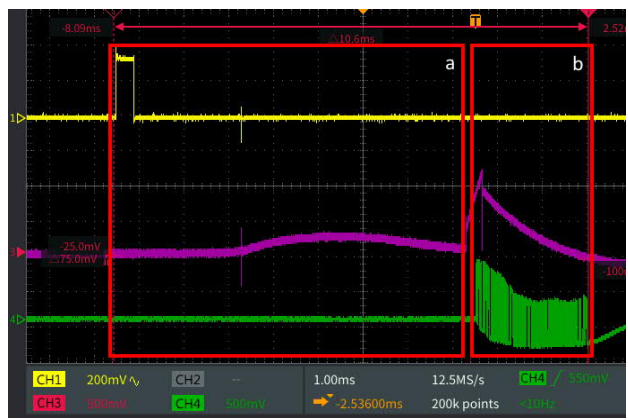

Figure 6. Minimum measured end-to-end latency.

Figures 7 and 8 shows the distribution of the latency for the 5G segment and for the overall system (5G link and VLC link), respectively. The latency was calculated over a transmission of 2250 packets. The most frequent latency for the 5G link is about 9.5 ms, while for the VLC link is about 2.5 ms. The 5G network latency time ranges from 2.4 to 29 ms, while the VLC network latency time varies from 2.4 to 3.1 ms. This time describes the data transmission time through the optical channel at the rate of 100 kbps, as defined by the IEEE 802.15.7 standard for outdoor applications. The processing time of the second Raspberry Pi 3 is highly stable, and it is equal to a few hundreds of µs.

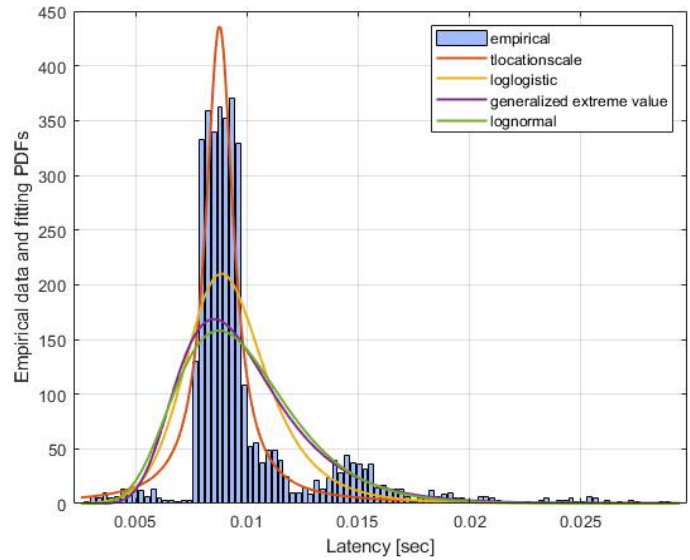

Figure 7. Empirical distribution of the latency of the 5G link only and PDFs of the fitting distributions.

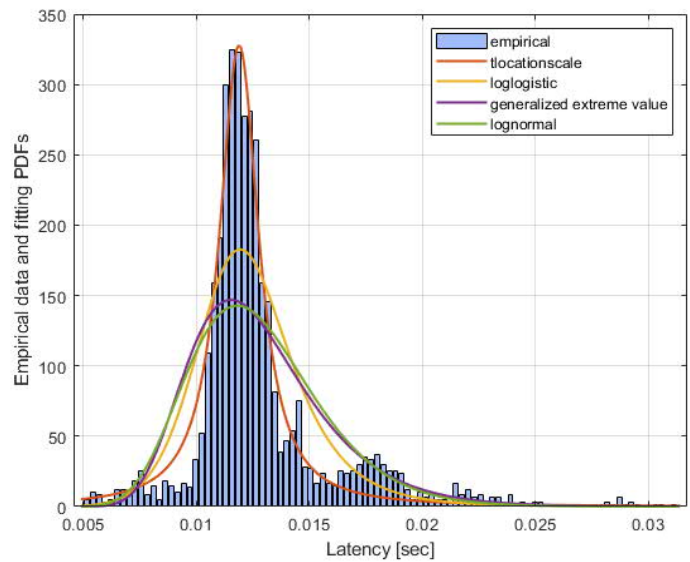

Figure 8. Empirical distribution of the latency of the overall system (5G link and VLC link) and PDFs of the fitting distributions.

The best fitting distribution of the latency for both the 5G link and overall system has been derived by using the Bayesian information criterion (BIC) [24]. Seventies different distributions have been evaluated as fitting models for the empirical data, and the distribution that minimizes the BIC was selected. For both the 5G link and the overall system, the distribution that best fits the empirical data is the t-location scale whose probability density function (PDF) $f(x|\mu, \sigma, v)$ is

$$f(x|\mu, \sigma, v) = \frac{\Gamma\left(\frac{v+1}{2}\right)}{\sigma\sqrt{v\pi}\Gamma\left(\frac{v}{2}\right)} \left[\frac{v+\left(\frac{x-\mu}{\sigma}\right)}{v}\right]^{-\left(\frac{v+1}{2}\right)} \qquad (2)$$

where $\Gamma(\bullet)$ is the gamma function, μ is the location parameter, $\sigma$ is the scale parameter, and $v$ is the shape parameter. The best fit t-location scale parameters [μ, $\sigma$, $v$] for the 5G link are [0.0088, 7.43 × 10$^{-4}$, 1.09], while for the overall system are [0.0119, 0.001, 1.253]. Table 2 summarizes the results.

Table 2. Best fitting distribution for 5G and VLC technologies.

| Technology | Latency: best fitting distribution | Distribution parameters [μ, σ, ν] |
|---|---|---|
| 5G | t-location scale | [0.0088, 7.43 × 10-4, 1.09] |
| VLC | t-location scale | [0.0119, 0.001, 1.253] |

Figures 9 and 10 shows the cumulative distribution function (CDF) of the four best fitting distribution models for the 5G link and for the overall system, respectively. As it can be seen, the CDF error is minimized by the t-location scale distribution in both cases.

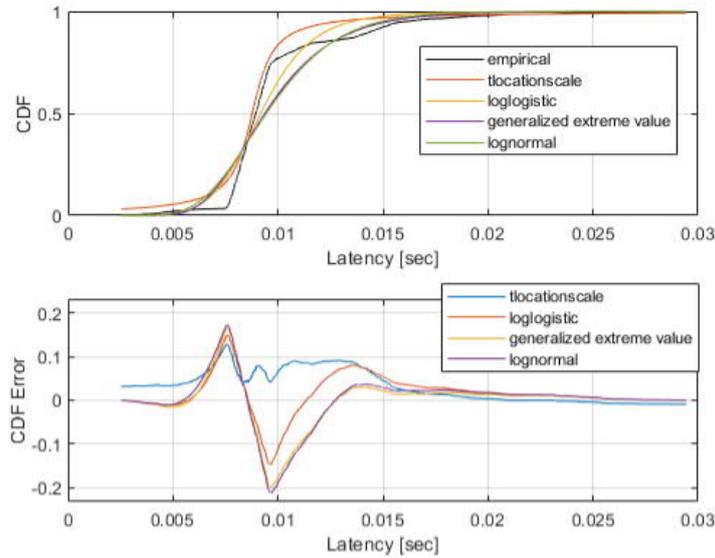

Figure 9. CDF and CDF error of the latency for empirical data and fitting distributions. 5G link only.

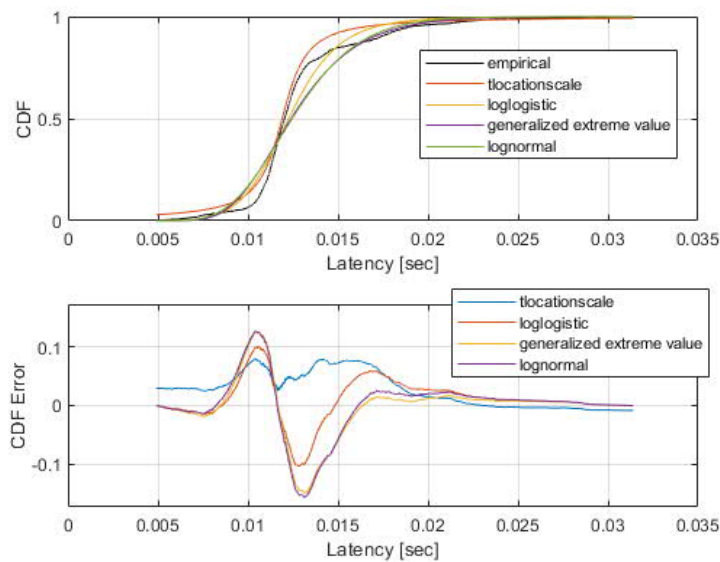

Figure 10. CDF and CDF error of the latency for empirical data and fitting distributions. Overall system (5G link and VLC link).

Although the NSA-5G network shows a very low latency compared to previous generations, we can assert that the longest part of the end-to-end latency time of the joint 5G-VLC network is introduced by the 5G part due to the Packet Error Rate (PER) optimization. The 5G network provider has implemented techniques to reduce the PER at the 5G receiver. This means that a redundancy of the transmitted packet is introduced, during the network set up phase, to obtain PER=$10^{-6}$ at the receiver. The VLC latency time could be even decreased by reducing the packet length, i.e., by optimizing the message length.

## 6. Conclusions

This paper introduced one of the first experiment on the joint use of a 5G and VLC network to provide information to cars. Data from road sensors has been collected by a 5G network and then sent to a VLC network for diffusion to cars through traffic-lights. In addition, data from on-demand emergency alarm has been implemented. The experiment aimed to measure the overall latency. The results showed that a total latency of about 12 ms can be reached. In particular, the most frequent latency for the 5G link was about 9.5 ms, while for the VLC link was about 2.5 ms. The main contribute is related to the 5G network part, that should be optimized, especially with the deployment of the SA 5G architecture.

**Author Contributions:** "Conceptualization, D. Marabissi and L. Mucchi; software, S. Caputo, T. Nawaz and M. Seminara; validation, T. Pecorella and J. Catani; data curation, S. Caputo and F. Nizzi; writing—original draft preparation, D. Marabissi and L. Mucchi; writing—review and editing, T. Pecorella and R. Fantacci; supervision, R. Fantacci; All authors have read and agreed to the published version of the manuscript.

**Funding:** This research was carried out as part of the 5G project, supported by the Italian Ministry of Economic Development.

**Conflicts of Interest:** The authors declare no conflict of interest.